\documentclass[preprint2]{aastex}

\shorttitle{Star-formation in low luminosity AGN}
\shortauthors{De Vries et al.}

\begin{document}
\title{Star-Formation in Low Radio Luminosity AGN from the Sloan Digital Sky Survey}

\author{W. H. de Vries, J. A. Hodge, R. H. Becker}
\affil{University of California, 1 Shields Ave, Davis, CA 95616}
\affil{Lawrence Livermore National Laboratory, L-413, Livermore, CA 94550}
\email{devries1@llnl.gov}

\author{R. L. White}
\affil{Space Telescope Science Institute, 3700 San Martin Drive,
Baltimore, MD 21218}

\and

\author{D. J. Helfand} 
\affil{Columbia Astrophysics Laboratory, Columbia University, 550 West
  120th Street, New York, NY 10027}

\begin{abstract}
 
We investigate faint radio emission from low- to high-luminosity
Active Galactic Nuclei (AGN) selected from the Sloan Digital Sky
Survey (SDSS). Their radio properties are inferred by co-adding large
ensembles of radio image cut-outs from the FIRST survey, as almost all
of the sources are individually undetected. We correlate the median
radio flux densities against a range of other sample properties,
including median values for redshift, [\ion{O}{3}] luminosity,
emission line ratios, and the strength of the 4000\AA\ break. We
detect a strong trend for sources that are actively undergoing
star-formation to have excess radio emission beyond the $\sim 10^{28}$
ergs s$^{-1}$ Hz$^{-1}$ level found for sources without any
discernible star-formation. Furthermore, this additional radio
emission correlates well with the strength of the 4000\AA\ break in
the optical spectrum, and may be used to assess the age of the
star-forming component.  We examine two subsamples, one containing the
systems with emission line ratios most like star-forming systems, and
one with the sources that have characteristic AGN ratios. This
division also separates the mechanism responsible for the radio
emission (star-formation vs. AGN). For both cases we find a strong,
almost identical, correlation between [\ion{O}{3}] and radio
luminosity, with the AGN sample extending toward lower, and the
star-formation sample toward higher luminosities. A clearer separation
between the two subsamples is seen as function of the central velocity
dispersion $\sigma$ of the host galaxy. For systems at similar
redshifts and values of $\sigma$, the star-formation subsample is
brighter than the AGN in the radio by an order of magnitude. This
underlines the notion that the radio emission in star-forming systems
can dominate the emission associated with the AGN.
  
\end{abstract}

\keywords{galaxies: active --- galaxies: starburst --- radio continuum: galaxies}

\section{Introduction}

The formation of the bulge component of galaxies and their central
black holes is now understood to be tightly linked
\citep[e.g.,][]{richstone98,kauffmann00,heckman04}. There are several
observational lines of evidence for this theoretical claim. For
instance, the central velocity dispersion of bulges is found not just
to correlate with the mass of the bulge, but also with the inferred
mass of the central black hole \citep[e.g.,][]{ferrarese00,gebhardt00}. 
Since the correlation between the mass of the central black hole and
the bulge exists over a large range of bulge masses, it is thought
that, above some lower mass limit, all bulges harbor massive central
black holes. Locally, accurate black hole mass measurements have been
made for our own Galaxy \citep[e.g.,][]{schoedel03,ghez05}, and the
Andromeda galaxy \citep[e.g.,][]{bender05}. Our own Galactic black
hole is rather dormant based on its relatively low X-ray emission
\citep[e.g.,][]{baganoff03,xu06}, but in general these massive black
holes are considered a key component of Active Galactic Nuclei (AGN),
in which accretion onto the black hole itself can produce large
amounts of energy efficiently \citep[e.g.,][]{lyndenbell69,pringle81}.

It is of interest, then, to study AGN properties over a large range of
luminosities, ranging from the AGN-emission-dominated quasars to
galaxies for which the presence of a weak AGN can be inferred through
unusual optical emission line ratios.  Kauffmann et al. (2003a,b)
studied the properties of the host galaxies of AGN, using large,
well-defined samples from the Sloan Digital Sky Survey \citep[SDSS,
e.g.,][]{york00,stoughton02}. The sheer size of these samples (a few
tens of thousands of galaxies) allows for very accurate measurements
of sample averages and their trends as function of optical continuum
and emission-line (mainly [\ion{O}{3}]) luminosities. One of the
conclusions from \citet[][hereafter K03a]{kauffmann03a} is that the
main difference between low- and high-luminosity AGN is the
significant presence of a young stellar population in the latter.

In this paper, we will concentrate on the radio emission properties of
a large sample of AGN, similarly selected from the SDSS survey
\citep[Data Release 4,][]{adelmanmccarthy06}. We use the latest AGN
compilation of Kauffmann et
al.\footnote{http://www.mpa-garching.mpg.de/SDSS/DR4} who used the BPT
\citep{baldwin81} diagram of emission line ratios to select for AGN
activity. The current version of this catalog contains 80,156
candidate objects which are also covered by the VLA\footnote{The Very
Large Array is an instrument of the National Radio Astronomy
Observatory, a facility of the National Science Foundation operated
under cooperative agreement by Associated Universities, Inc.} FIRST
radio survey \citep{becker95}. In addition, we selected 14,165 quasars
from the DR4 release for which we have FIRST images and [\ion{O}{3}]
line coverage (limiting the quasar redshifts to $<0.8$). These quasars
will form the high-luminosity portion of our sample.

We focus our attention on the radio emission of the AGN for several
reasons. First, almost all AGN are thought to produce radio flux at
some level; even the Galactic black hole, currently in a ``quiescent''
phase, emits in the radio. Some AGN are extremely luminous in the
radio (compared to their optical output) -- the so-called radio-loud
objects, although we are not considering these specifically, as our
average AGN is radio-quiet by a wide margin. Secondly, radio emission
is unaffected by intrinsic absorption within each source. It therefore
provides a more or less accurate measure of AGN radio output, with the
sole caveat being relativistic beaming effects. Since none of our
sources are classified as Blazars for which beaming effects can be
large, we assume this to play a minor role. Finally, the ready
availability of a large-area, high-resolution radio survey in FIRST
provides radio information on the large numbers of AGN in its 9033
deg$^2$ coverage.

There is, however, one apparent problem: almost none of the sources in
the SDSS sample are detected by the FIRST survey, with its flux
density threshold at 1.4GHz of 1.0 mJy (5$\sigma$). Out of 80\,154
AGN, only 5\,875 (7.3\%) are detected directly by the FIRST
survey. The relevant numbers for the quasars are only slightly better:
1\,493 out of 14\,165 (10.5\%). Clearly by limiting the sample to the
brightest sources, one is ignoring the bulk ($\sim 90$\%) of the
population.  Fortunately, we can stack radio images, and detect the
mean and median peak flux densities of ensembles of undetected
sources. This method is discussed in depth by \citet[][hereafter
Paper~I]{white06}; we summarize the method in \S~\ref{stacking}.
Given our large sample size, we are able to detect not just the
(median) ensemble flux density values, but also discern small trends
in those values. For our typical stack sample size of 5000 sources, we
attain an rms noise of 2.6$\mu$Jy, and are therefore very sensitive to
small changes in the sample median values as we vary other parameters
such as the redshift distribution and the [\ion{O}{3}] luminosity
distribution. We only have a single sample, so the way we investigate
the varying dependencies is by re-grouping the stacks according to the
quantity of interest. For instance, if we order the sample by redshift
and generate 15 adjacent bins (5343 sources per bin), the resulting
radio stacks will be different from the 15 bins sorted by [\ion{O}{3}]
luminosity; each plot in this paper contains the exact same data, but
arranged differently.

The typical median radio flux densities for these sources is found to
be on the order of 50 to 100 $\mu$Jy, well below the FIRST threshold,
but clearly within the capabilities of the VLA. Indeed, various very
deep, small-scale radio surveys exist \citep[e.g.,][]{hopkins98,
richards99, devries02}. At these low flux densities, star-formation
could account for a large fraction of the radio emission in our
objects, since flux density levels of 50 $\mu$Jy are well within range
of starforming galaxies (without AGN) at similar redshifts \citep[see,
e.g.,][]{windhorst99, fomalont02}. We cannot, therefore, assign all of
the radio emission to the AGN, as some fraction of the emission may be
due to star-formation.

The paper is organized as follows. In \S~\ref{stacking} we summarize
the stacking method as presented in Paper~I. The next section
describes the properties of the complete sample of 80,156 sources. We
then use emission line ratios as well as the strength of the 4000\AA\
break to isolate two subsamples of 26,715 sources each
(\S~\ref{BPTsection}). The first subsample has properties most
consistent with the presence of an AGN (the ``pure'' AGN sample); the
second has star-formation characteristics while still meeting the AGN
selection criterion of K03a. Section \ref{agnresults} details the
differences and similarities of these subsamples, which are discussed
in further in \S~\ref{discussion}.

\section{FIRST image Stacking Technique}\label{stacking}

Paper~I demonstrates that it is possible to measure the mean and
median radio flux density values of distributions of sources, even
though individual sources fall far below the detection threshold of
the FIRST survey. Provided that the sample is large enough, one can
attain rms noise values in the radio sky well below the canonical
FIRST value of 0.15 mJy. Actual snap-shot stacking experiments of
blank pieces of sky\footnote{No pieces of the sky are truly blank, but
will consist of faint, unresolved background objects. Our stacking
technique is not sensitive enough to pick up this signal: we measured
a background signal consistent with 0$\mu$Jy to within the 0.9$\mu$Jy
rms noise (for 80\,154 empty patches).} conform to the expected
$1/\sqrt{N}$ behavior in the pixel statistics, down to better than
1$\mu$Jy. Based on these numbers, it is clear that one can detect
stacked point-source mean flux densities of a few tens of $\mu$Jy with
high fidelity.

As outlined in Paper~I, we prefer to use the median value of the
distributions over the mean. The latter quantity is rather easily
affected by outliers with large flux densities (the ones which are
actually above the FIRST detection threshold). Simply removing the
sources above the threshold from the sample to arrive at the mean of
the undetected sources is not robust, as a small change in the cut-off
flux density (e.g., from 1.0 to 0.9 mJy) results in a significant
change in the mean flux value.

The second concern addressed in Paper~I is the calibration of the
stacking procedure for effects introduced by the radio data analysis
in general and the 'CLEAN' algorithm in particular. Analysis of both
stacked artificial sources, and actual sub-threshold sources detected
in full-synthesis, deeper radio imaging shows that one does not
recover fully the flux that went into the stack. These results are
illustrated in Fig.~2 of Paper~I; the correction for this
``snapshot-bias" for sub-threshold sources is given by $S_{p,corr} =
1.40 S_p$, where $S_p$ is the median peak flux density value. All of
our stacked peak flux density measurements have been corrected by this
factor.

 \subsection{Stacking luminosity images}

Since most of our subsequent discussion deals with radio luminosities,
and not radio flux densities, we have to consider whether there are
any distribution peculiarities that might affect these quantities. We
do not know the flux density distribution of sub-threshold sources,
nor the nature of the sources that make up this population. Below flux
densities of a few mJy, the composition of the radio source population
changes from AGN-dominated to star-formation dominated
\citep[e.g.,][]{windhorst99}.

There are two ways of calculating the median radio luminosity. One is
simply to stack the cut-outs (with pixels in units of Jy), and use the
median peak flux density and median redshift (of the sources that were
used in the stack) to calculate the median radio luminosity. The tacit
assumption here is that those two median values actually describe the
same ``median object''. The alternative approach is to convert each
snapshot into a luminosity image using the redshift of the individual
source, and then stack these images instead. Each pixel in a snapshot
image is converted from flux density to luminosity using the following
cosmological parameters: $H_{\rm o} = 70$ km s$^{-1}$ Mpc$^{-1}$,
$\Omega_\Lambda = 0.7$, and $\Omega_{\rm M} = 0.3$.

It turns out that the results are not exactly the same.  Table~2
(columns 4 and 5) lists the radio luminosities derived using both
methods for the sample sorted in D$_{\rm n}$(4000) strength, showing
typical differences of less than 25\%. For the sake of consistency, we
will use the luminosity-stacked values instead of the flux
density-stacked ones for the remainder of the paper.

\section{Star-formation vs. AGN activity}\label{BPTsection}


\begin{figure}[t]
\plotone{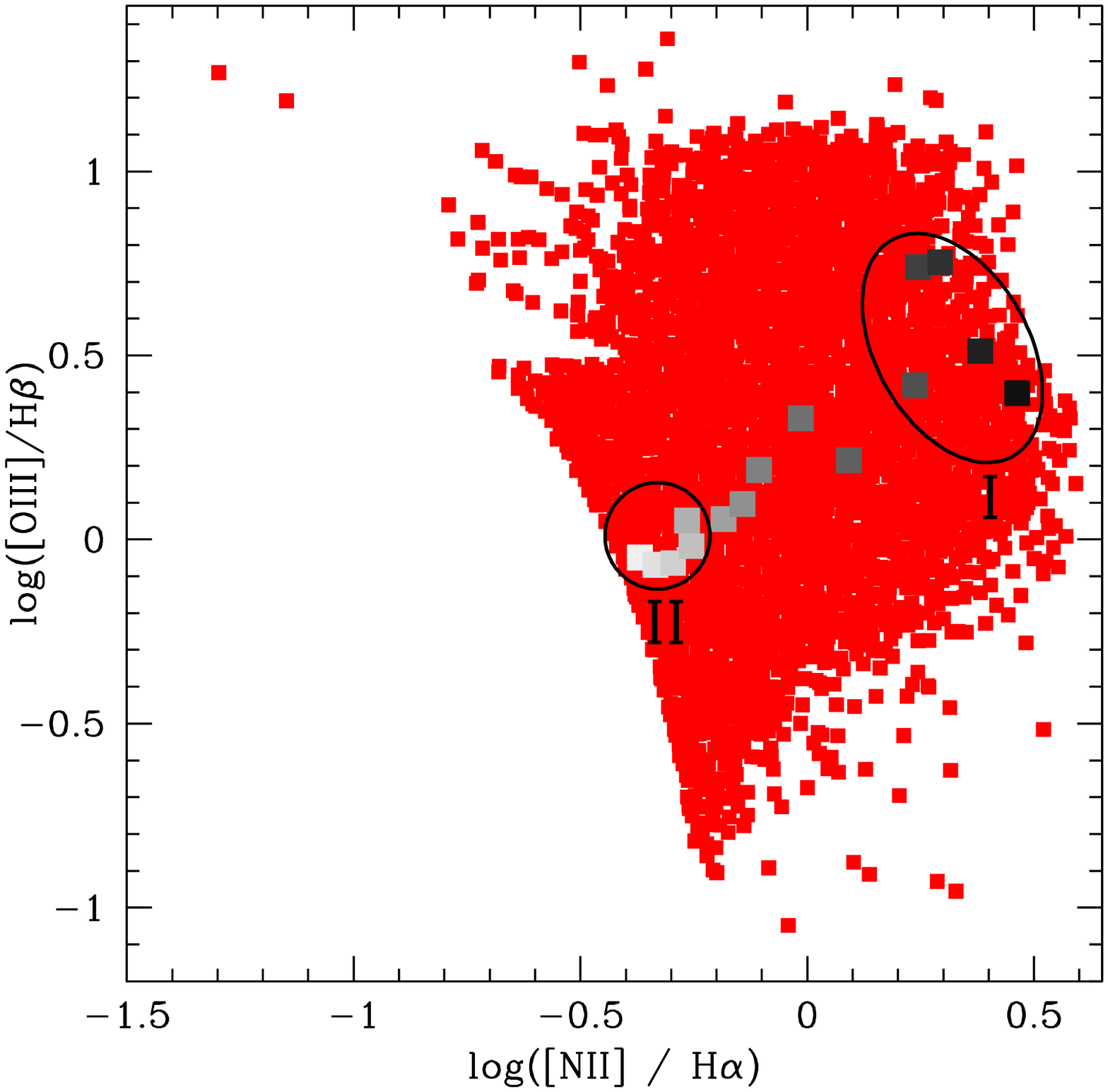}
\caption{Emission line diagnostic plot \citep[BPT,][]{baldwin81} which
  distinguishes between star-formation-dominated and AGN-dominated
  systems. The sharp cutoff reflects the selection boundary of
  \citet{kauffmann03a}. Points to the left of this cutoff are
  considered pure star-formation without an AGN contribution. Pure AGN
  ratios are only expected in the extreme upper right hand corner, so
  most objects are mixes. The greyscale squares illustrate the
  progression of low-strength D$_{\rm n}$(4000) systems (i.e., most
  actively star forming) to large break systems (light to dark grey
  respectively). The circled groups, I and II, are defined in
  Fig.~\ref{d4ktrend}.}
\label{BPTplot}
\end{figure}

\begin{figure}[t]
\plotone{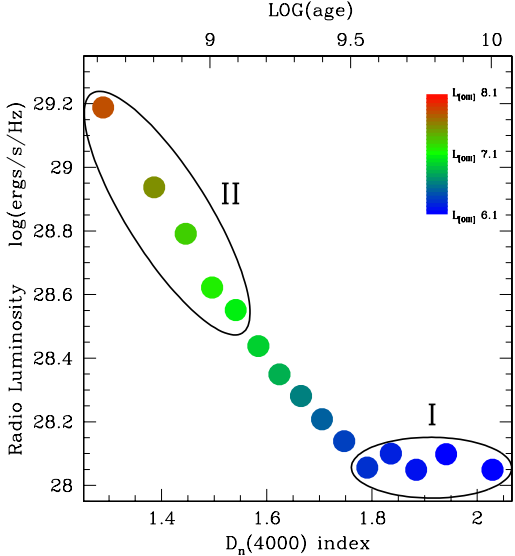}
\caption{Radio Luminosity as function of D$_{\rm n}$(4000)
  strength. The bins, each containing about 5000 sources, are sorted
  in order of increasing D$_{\rm n}$(4000) strength (from left to
  right), and are coded for [\ion{O}{3}] luminosity. The D$_{\rm n}$
  index is not very dependent on redshift; the median redshifts range
  from 0.12 down to 0.07 (from left to right in the plot). The age
  indicated on the top of the plot is the inferred age since the last
  burst of star-formation (in years), based on a linear conversion
  between D$_{\rm n}$(4000) and age (see text). The sub-samples marked
  I and II are described in the text.}
\label{d4ktrend}
\end{figure}

Our data sample was initially selected based on source locations in
the BPT plot (Kauffmann et al. 2003a, see also Kewley et al. 2006).
The ratios of the narrow-lines of [\ion{N}{2}] and [\ion{O}{3}] over
the permitted Balmer lines H$\alpha$ and $H\beta$, respectively, serve
as an excellent proxy for the amount of star-formation in an object
\citep[e.g.,][]{osterbrock89}. Our galaxies with AGN cover a large
range in the BPT plot, and are thus affected by star-formation to
varying degrees; the objects found close to the sample cut-off arc,
toward the left of the distribution (see Fig.~\ref{BPTplot}), have
optical spectra dominated by emission line ratios commonly associated
with ongoing star-formation. It should also be noted that Kauffmann et
al. used a more liberal cut for their AGN classification than for
instance, \citet{kewley01}, resulting in a fraction of the sources
being more H II region-like than AGN-like. These sources are
classified as ``composite'' in the related \citet{brinchmann04} paper.

We use the 4000\AA\ break strength index D$_{\rm n}$(4000) as defined
by \citet{kauffmann03b}. This index can be parameterized as the age
since the last (instantaneous) episode of star-formation (see Fig.~2
of Kauffmann et al. 2003b), in the sense that smaller break strengths
correspond to more recent star-formation\footnote{It should be noted
that Kauffmann et al. (2003a) found that the AGN contribution to the
blue continuum, which would give rise to an overestimate of the
star-formation rate using this method, is small enough to be ignored
for these objects.}. The ranges of the D$_{\rm n}$(4000) under
consideration for our sources roughly correspond to ages since
star-formation between 10$^{8.6-10.0}$ years. The short end of this
range is comparable to typical galaxy merger timescales
\citep[e.g.,][]{springel05}, whereas $\sim10^{10}$ years at $z \sim
0.2$ is close to the age of the universe and therefore represents the
initial burst of star-formation. Since we are not going to distinguish
between starbursts of different metallicities, we are approximating
the relation between age and D$_{\rm n}$(4000) by a straight line over
the D$_{\rm n}$(4000) range 1.3 to 2.0 (the relevant range for our
sample): $\log{(\rm age)} = 6.21 + 1.87 \, {\rm D}_{\rm n}(4000)$,
with age given in years.

The results for median radio luminosity as function of the strength of
the 4000\AA\ break are plotted in Fig.~\ref{d4ktrend}. It is clear
that the brightest radio emission is associated with the smallest
breaks (i.e., the most recent episodes of star-formation). The
approximate time since the last epoch of star-formation is given
across the top of the plot. Above a D$_{\rm n}$(4000) index of about
1.8, there is no change in radio luminosity, which levels off at about
$10^{28}$ ergs s$^{-1}$ Hz$^{-1}$. This might reflect the absence of
any contribution to the radio emission from star-formation. The color
coding indicates the [\ion{O}{3}] luminosity, which is also seen to
decline steeply as D$_{\rm n}$(4000) increases.

\subsection{Line emission diagnostics}

\begin{figure}[t]
\plotone{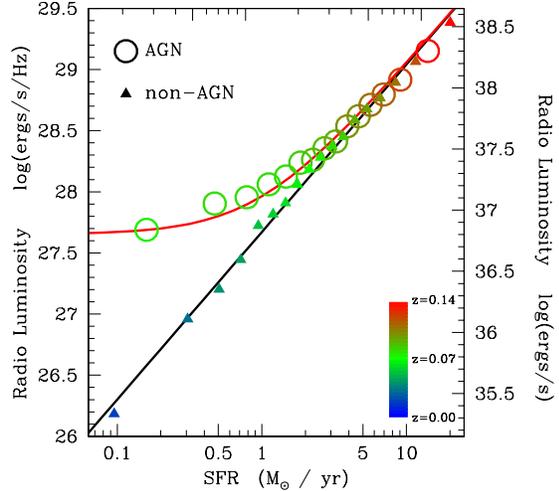}
\caption{Correlation between fiber aperture-corrected star-formation
  rate, based on the strength of the D$_{\rm n}$(4000) break
  \citep{brinchmann04}, and the radio luminosity. The circles
  represent our sample, and the small triangles are star-forming
  galaxies without an AGN (taken from Brinchmann et al.). The
  least-squares fit to the non-AGN data points (black line) is given
  by: $\log(L_{\rm R}^{\rm non-AGN}) = (1.37\pm0.02) \log{({\rm SFR})}
  + (27.67\pm0.01)$. The red line represents the sum of the non-AGN
  component $L_{\rm R}^{\rm non-AGN}$ and a constant
  $4.5\times10^{27}$ ergs s$^{-1}$ Hz$^{-1}$ due to the AGN.}
\label{sfrtrend}
\end{figure}

As a double check on the level of star-formation, we can see where the
D$_{\rm n}$(4000)-sorted data points from Fig.~\ref{d4ktrend} land on
the BPT plot. In other words, do the sources with the low D$_{\rm
n}$(4000) values have emission line ratios that reflect a higher
incidence of star-formation? The results are plotted in
Fig.~\ref{BPTplot} as a grayscale sequence of squares, running from
light-gray to black for the small-to-large valued D$_{\rm n}$(4000)
systems. It is clear that these two quantities do correlate rather
well for our objects, and trace out a sequence of decreasing
star-formation versus increasing 4000\AA\ break strength. The gray
squares in the top right corner have emission line ratios completely
consistent with pure AGN emission, whereas the squares to the left are
significantly affected by star-formation.

The combination of Figs.~\ref{BPTplot} and \ref{d4ktrend} leads us to
suggest that for the ``pure'' AGN cases as indicated by large 4000\AA\
breaks, the radio luminosity is strictly due to the AGN with average
luminosities near $10^{28}$ ergs s$^{-1}$ Hz$^{-1}$. The additional
radio luminosity we see for other sources can be attributed completely
to ongoing star-formation. This star-forming component can be an order
of magnitude more luminous than the radio component due to the AGN.

\citet{brinchmann04} use the strength of the 4000\AA\ break to
calculate the inferred star-formation rate (their sample not only
encompasses our AGN sample, but also non-AGN which land to the left of
the distribution in Fig.~\ref{BPTplot}). They corrected for the
limited angular extent of the SDSS fiber which typically does not
contain all of the light of the galaxy. We use these corrected values
to sort our sample in ascending order of star-formation rate. The
median stacked radio luminosities for 15 bins of $\sim5000$ sources
each are calculated and plotted against median star-formation rate in
Fig.~\ref{sfrtrend}. The strong correlation again underlines the
importance of the star-formation component to the overall radio
luminosity, even for systems containing AGN. The $10^{28}$ ergs
s$^{-1}$ Hz$^{-1}$ ``ground-state'' radio luminosity from
Fig.~\ref{d4ktrend} corresponds, based on this plot, to a
star-formation rate of $\sim1$ M$_\odot$ yr$^{-1}$, or less, whereas
the largest radio luminosities from Fig.~\ref{d4ktrend} have rates
exceeding 10 M$_\odot$ yr$^{-1}$.

The non-AGN portion of the Brinchmann et al. sample, once sorted by
star-formation rate, exhibit a very tight correlation with median
radio luminosity (Fig.~\ref{sfrtrend}). The least squares fit is given
by $\log(L_{\rm R}^{\rm non-AGN}) = (1.37\pm0.02) \log{({\rm SFR})} +
(27.67\pm0.01)$. This slope of 1.37 is almost identical to that of
\citet{bell03} who find a slope of 1.30 for radio luminosities less
than $6.4\times 10^{28}$ ergs s$^{-1}$ Hz$^{-1}$. The Bell results,
however, show a different normalization, suggesting they underestimate
the amount of star-formation for a given radio luminosity by a factor
of 2.

Even more remarkable than the tight correlation between estimated
star-formation rate and mean radio luminosity over two orders of
magnitude is that the AGN contribution to the radio luminosity can be
represented by a constant ($4.5\times10^{27}$ ergs s$^{-1}$ Hz$^{-1}$)
which, once added to the star-formation estimate (black line), gives
us the red line. This single constant implies that the AGN output (in
the radio) is more or less independent of the star-formation level of
the host galaxy. It is also a better measure of this AGN
"ground-state" than the $10^{28}$ value derived from
Fig.~\ref{d4ktrend}.

In summary, since the star-formation rates are based on the D$_{\rm
n}$(4000) values, we will use the latter to isolate two subsamples;
one that presumably has no, or very little, ongoing star-formation
(labeled I in Figs.~\ref{BPTplot} and \ref{d4ktrend}), and one that
contains the systems with the most ongoing star-formation (labeled
II). Each one of these subsamples contain 26,715 sources.

\section{A closer look at the ``pure'' AGN subsample}\label{agnresults}

\begin{figure}[t]
\plotone{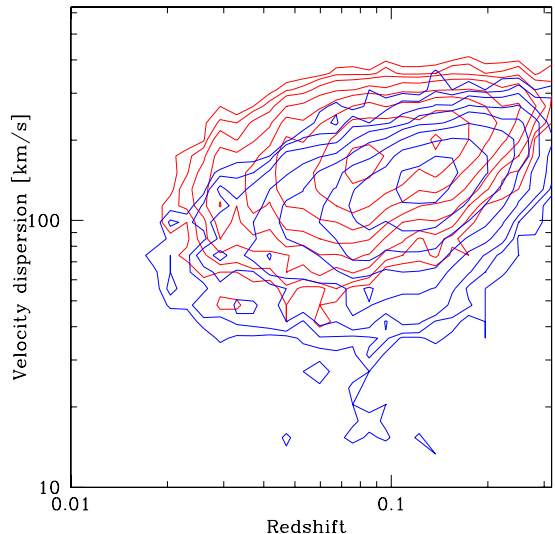}
\caption{Distribution densities for the 5 lowest (red) and 5 highest
  (blue) radio luminosity bins of Fig.~\ref{d4ktrend}. While the
  redshift coverage is comparable, there is a clear dependency on the
  strength of the D$_{\rm n}$(4000) index in the sense that it is
  easier for a given burst of star-formation to ``fill in'' the
  4000\AA\ break in a smaller, low $\sigma$ system. The density
  contours are given by $2^n$, with $n=2,3,4, \cdots$.}
\label{compzlvd}
\end{figure}

We designate the five rightmost data points in Figs~\ref{BPTplot} and
\ref{d4ktrend} as representing the most AGN-like sources. Their
emission line ratios are the least star-formation-like, and their
median [\ion{O}{3}] and [\ion{O}{2}] luminosities are among the lowest
in our sample (see Table~2).

Figure~\ref{compzlvd} shows the distribution densities in redshift
versus velocity dispersion space for the large 4000\AA\ break, low
star-formation sources (red contours), compared to the sources with
the smallest 4000\AA\ break and the highest levels of star-formation
(blue contours). The redshift distribution for each subset is
comparable, whereas there is a clear offset in velocity
dispersion. This is most likely due to the way the subsets have been
selected. For a given burst of star-formation with some upper limit to
its optical luminosity, it will appear more prominent (with a
consequently lower value of the D$_{\rm n}$(4000) index) in smaller
galaxies than in large systems. Since the mass and size of the galaxy
spheroid scales with its central velocity dispersion
\citep[e.g.,][]{gebhardt00,ferrarese00}, we will be biased toward
finding more star-formation among the lower $\sigma$ systems.

\subsection{Radio Luminosity and redshift dependence}\label{radioz}


\begin{figure}[t]
\plotone{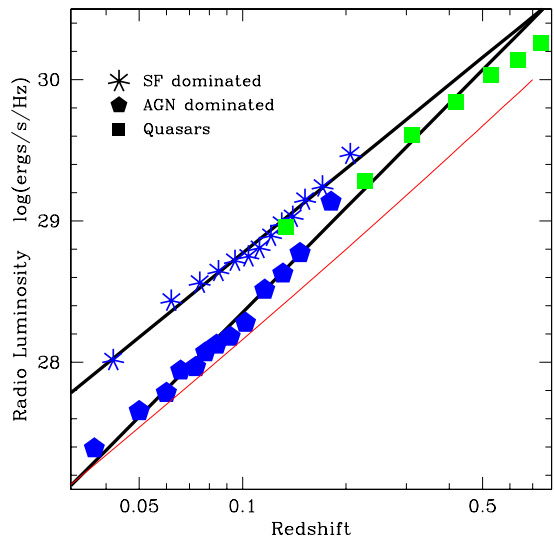}
\caption{Radio luminosity as function of redshift. The narrow-line AGN
  sample is separated into AGN dominated (I, pentagons) and
  star-formation dominated (II, stars) parts. In addition, 7
  low-redshift quasar bins (squares) are overplotted. The least
  squares fits are: $\log(L_{\rm R}) = (1.98\pm0.07) \log{(z)} +
  (30.75\pm0.07)$, and $\log(L_{\rm R}) = (2.45\pm0.11) \log{(z)} +
  (30.81\pm0.12)$ for subsamples I and II respectively. The red solid
  line indicates the radio luminosity of a constant 60 $\mu$Jy
  source.}
\label{ztrendRadio}
\end{figure}

\begin{figure}[t]
\plotone{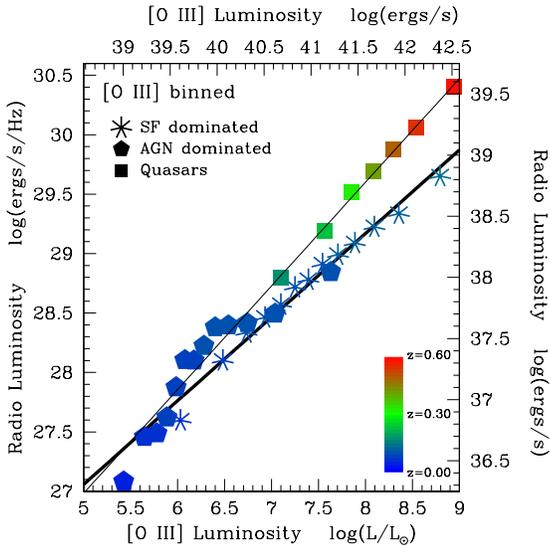}
\caption{Correlation of median values for radio luminosity and
  [\ion{O}{3}] luminosity. The color indicates the median redshift for
  a particular bin. The 7 squares are broad-line AGN / Quasars. The
  units on the top and right hand side are converted from the measured
  values by the following constants: $\log{(1.4\times10^9)}$ Hz, and
  $\log{(3.826\times10^{33})}$ ergs/s for the x-axis (observing
  frequency) and y-axis (bolometric luminosity Sun) respectively. Note
  that for all these points the radio luminosity is at least 3 orders
  of magnitude lower than typical radio-loud objects. Symbols are the
  same as in Fig.~\ref{ztrendRadio}.}
\label{zandotrend}
\end{figure}

Luminosity, as in all flux-limited samples, is closely correlated with
redshift, resulting in an apparent correlation that is dominated by
the lack of bright sources locally, and faint sources at higher
redshifts (due to the detection limit). This is illustrated in
Fig.~\ref{ztrendRadio}. Both subsamples I and II were sorted in
redshift, and divided up into 13 bins of 2056 sources each. The
``pure'' AGN-like bins are plotted as pentagons, and the sources most
affected by star-formation as stars. In addition, we plotted 7 bins of
2025 quasars each, depicted as green squares.

It is immediately clear that the median radio luminosities of the
quasars are directly comparable to the narrow-line AGN values. There
is no a-priori reason for the quasar radio luminosity medians not to
be much brighter than the galaxies, but their values are comparable at
the same redshift. The quasar radio luminosities are slightly less
than the star-formation values, and slightly more than the AGN
subsample values (at the same redshift).  On the other hand, quasars
{\it are} consistently brighter in [\ion{O}{3}] luminosity than the
narrow-line AGN at similar redshifts (see Fig.~\ref{zandotrend}).

The fact that the median radio luminosity is the same would suggest
that we are observing an isotropic emission component, and not
orientation / obscuration-dependent quantities like emission lines and
the presence of broad permitted lines. The median radio luminosity
therefore underlines the validity of the standard AGN unification
model \citep[e.g.,][]{barthel89,antonucci93}.

Another result based on Fig.~\ref{ztrendRadio} is that the
star-formation-dominated galaxies are brighter in the radio than the
pure AGN galaxies, albeit that the difference becomes smaller at
redshifts beyond 0.1 (where both samples I and II merge into the low
end of the quasar trend). The more or less constant offset of a factor
$\sim 3$ (0.5 in log) between the two subsamples (below $z \sim 0.1$)
suggests that the average radio luminosity due to the AGN accounts for
at most 25\%\ of the total radio emission during an ongoing phase of
star-formation.

\subsection{[\ion{O}{3}] luminosity as AGN activity indicator}\label{o3radio}

The [\ion{O}{3}] emission-line luminosity offers another way of
gauging the AGN output. \citet{kauffmann03a} estimate that perhaps as
little as 7\%\ of the [\ion{O}{3}] line emission can be attributed to
star-formation in their AGN composite spectrum.  This means that, even
though both the radio and [\ion{O}{3}] luminosities depend on
redshift, their correlation should be relatively free of redshift bias
(by dropping out of the ratio).

For this purpose, we first order the sample in [\ion{O}{3}]
luminosity, and then stack the cutouts to calculate the median radio
luminosity. The [\ion{O}{3}] luminosities have been corrected for
extinction using the H$\alpha$ / H$\beta$ emission line ratio, under
the assumption that the [\ion{O}{3}] emission is co-spatial (see
K03a). The [\ion{O}{3}] sorting and stacking also helps in minimizing
the redshift differences among the bins.  Ideally, one would like to
have similar source redshift distributions for each [\ion{O}{3}]
luminosity bin, which would in effect isolate one from the other.

The results are presented in Fig.~\ref{zandotrend}.  We see a strong
correlation between [\ion{O}{3}] and radio luminosity. For comparison,
we also included seven bins of low-redshift quasars. The spectral
coverage of the SDSS is such that the [\ion{O}{3}] emission line is
included in the spectrum up to redshifts of $\sim0.8$. This results in
a sample of 14,165 DR4 quasars that have $z < 0.8$ and are covered by
the FIRST survey. We split the sample into seven equal-sized bins,
sorted in [\ion{O}{3}] luminosity. Since we are looking at presumably
unobscured quasars, we do {\it not} correct the [\ion{O}{3}]
luminosities for intrinsic extinction\footnote{We know that the main
sample of galaxies has to have some extinction towards the AGN (they
would otherwise appear as type~1 Seyfert objects), so correcting for
the [\ion{O}{3}] luminosity (mainly due to the AGN) makes sense. But
since the quasars {\it are} type~1 objects, we will assume that the
corrections for the quasars are considerably smaller than the typical
factor $\sim10$ corrections in the median [\ion{O}{3}] luminosity for
the galaxies.}.

The two solid lines are least squares fits to the quasars and the
star-formation subsample. The first ten data points of the AGN
subsample (pentagons) are more or less consistent with the quasar
relation, though we do note that there are excursions along the quasar
fit. This may have to do with the extinction correction being less
than ideal for the AGN subsample.  The star-formation subsample does
not exhibit similar behavior. The fits are:

\begin{eqnarray}
L_{r,{\rm AGN}} = (0.87\pm0.02) L_{\rm [O III]} + (2.5\pm0.8) \nonumber\\
L_{r,{\rm SF}} = (0.70\pm0.03) L_{\rm [O III]} + (9.1\pm1.2)
\end{eqnarray}

\noindent with all the luminosities in units of log(ergs/s). It should
be noted that the origin of the radio emission is different for the
two subsamples. For the AGN- dominated sample the radio emission is
largely due to the AGN (though the star-formation component dominates
at the highest luminosities, see Fig.~\ref{sfrtrend}), whereas the AGN
component is only a small contributor for the star-formation
subsample. As such, it is clear that the differences in the
[\ion{O}{3}] -- radio luminosity relations cannot be explained by a
simple translation along either the [\ion{O}{3}] or the radio
luminosity axes. From Fig.~\ref{ztrendRadio} we already know that the
starforming subsample is brighter in the radio than either the AGN or
quasar samples (at the same redshifts). Therefore, the correct
conclusions from Fig.~\ref{zandotrend} are: the starforming subsample
is both brighter in [\ion{O}{3}] {\it and} in radio luminosity than
the AGN subsample at the same redshift, and they have a different
slope. Also, the AGN subsample has a similar slope to the quasar
sample, which is what one would expect if the AGN are obscured
quasars.
   
\subsection{Velocity Dispersion}

\begin{figure}[t]
\epsscale{0.9} 
\plotone{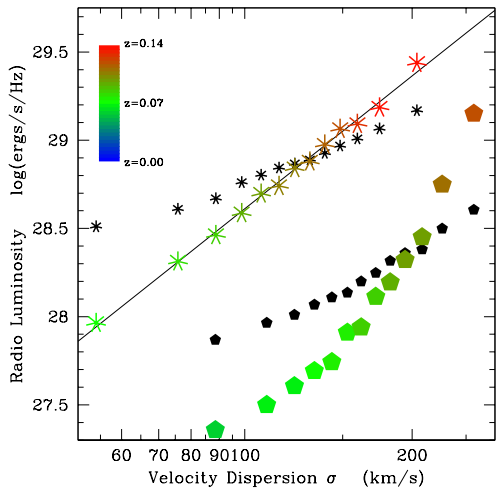}
\caption{Median radio luminosity as function of central velocity
  dispersion $\sigma$ of the host galaxy, for the AGN dominated (I,
  pentagons) and star-formation dominated (II, stars) subsamples. The
  spread in median redshift is indicated by the color coding. The
  close correlation between the median values of the velocity
  dispersion and the radio luminosity is apparent, especially for the
  star-formation sample.  The small symbols indicate where the colored
  dots would lie if we use its median redshift and the redshift --
  radio luminosity correlation of Fig.~\ref{ztrendRadio}. Note that
  the offsets are along the y-axis. Low $\sigma$ galaxies are
  underluminous in the radio compared to the average redshift trend,
  and the most massive galaxies are overluminous in the radio (this
  holds true for both subsamples).}
\label{lvdtrend}
\end{figure}

The central velocity dispersions of galaxies have been found to
correlate tightly with the inferred masses of their central black
holes \citep[e.g.,][]{ferrarese00,gebhardt00}. The latter quantity
also scales with the mass of the stellar bulge. There is some evidence
that the most massive bulge systems are more likely to be radio loud
compared to less massive systems \citep[e.g.,][]{laor00,lacy01}, but
in our cases we derive median radio luminosities well below what one
would consider radio-loud objects. As such, our analysis pertains more
to ordinary systems covering a range of AGN indicators (see
\S~\ref{BPTsection}).

The results are shown in Fig.~\ref{lvdtrend}. It is clear that there
is a trend for the most massive systems (highest values of $\sigma$)
to have the highest radio luminosities. This should not come as a
surprise, based on the correlations between black hole mass and bulge
mass. As seen in \S~\ref{radioz}, the median radio luminosity is
apparently not very much affected by orientation effects (since the
radio luminosities are comparable for quasars and AGN to within a
factor of two at the same redshift), and provides a rather clean
measure of the AGN output (at least for the ``pure'' AGN
subsample). This then would presumably scale nicely with the mass of
the central black hole, giving rise to the tight correlation between
median radio luminosity and velocity dispersion (see
Fig.~\ref{lvdtrend}, pentagon symbols).  The least-squares fit to the
star-formation subsample (star-shaped symbols) is given by:

\begin{equation}
L_{r,{\rm SF}} = (2.50\pm0.05) \log{(\sigma)} + (23.61\pm0.10).
\end{equation}

\noindent The AGN-dominated sample does not allow for a straight line
(in log-log) fit, which may indicate other dependencies beyond a
simple velocity dispersion / galaxy mass scaling.

One thing we need to check though, is the effect the small range in
redshift may have on radio luminosity. Based on their tight
correlation in Fig.~\ref{ztrendRadio}, a small change in median
redshift may account for most of the trend we see in
Fig.~\ref{lvdtrend}. To this end, we calculate the median redshift of
the sources used for each velocity dispersion bin, and applied the
redshift -- radio luminosity fits from Fig.~\ref{ztrendRadio} to infer
the radio luminosity that way. These values are overplotted as the
smaller black symbols (so each $\sigma$ bin has two radio luminosities
associated with it). Now, if the large and small symbols would closely
track each other, then the correlation between velocity dispersion and
radio luminosity is spurious due the sample being flux-limited.
However, since this is not the case, the radio luminosity -- velocity
dispersion correlation is real.

Note that the offset in radio luminosity between the star-formation
dominated and ``pure'' AGN subsamples is even larger than in
Fig.~\ref{ztrendRadio}, at about an order of magnitude compared to a
factor of about 3 (for $z < 0.1$). Clearly for some objects at similar
redshifts and similar mass (velocity dispersion), star-formation can
account for 90\%\ of the total radio luminosity. This would also
suggest that the correlation for the star-formation subsample is
driven by the bulge mass -- $\sigma$ relation (higher $\sigma$ equals
a more massive galaxy, which translates into a higher absolute amount
of star-formation for a given D$_{\rm n}$(4000) value). The radio
luminosity for the ``pure'' AGN subsample (pentagons in
Fig.~\ref{lvdtrend}) on the other hand, scales with the mass of the
central black hole, which itself scales with the bulge mass (which is
directly related to the velocity dispersion).

\section{Discussion and Summary}\label{discussion}

\begin{figure}[t]
\plotone{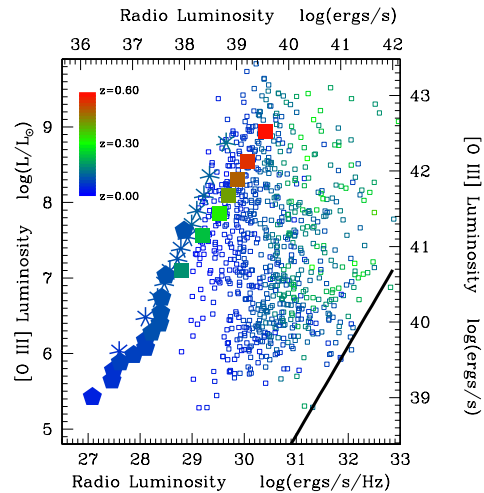}
\caption{Comparison of the correlation between [\ion{O}{3}] and radio
  luminosities (see Fig.~\ref{zandotrend}, note that the axes are
  switched) to values from the literature. The sample of Best et
  al. (2005) contains SDSS objects with a radio flux of $> 5$mJy and
  are represented by the small open squares (color coded with
  redshift). The offsets in radio luminosity between our median values
  and theirs can exceed 3 orders of magnitude. It should be noted that
  our sample of 80,156 includes the Best et al. sample, underlining
  the severe bias one introduces by selecting the brightest radio
  sources. The solid line is the fit from \citet{wills04} on local
  low-luminosity AGN.}
\label{bestcomp}
\end{figure}

We note the correlations of the median values of the [\ion{O}{3}] and
radio luminosities in \S~\ref{o3radio} and Fig.~\ref{zandotrend}, in
particular the similar ratios for the AGN, star-formation, and quasar
subsamples. Given that the mechanisms responsible for the radio
emission are thought to be different between the AGN and
star-formation subsamples, this comes as a surprise. Our work is not
the first that that touches on this subject. For instance,
\citet{wills04} worked on a sample of nearby ($z < 0.05$),
low-luminosity AGN taken from a 2Jy radio sample. They found an
[\ion{O}{3}] -- radio luminosity slope of 0.85, based on a sample of a
few dozen sources\footnote{We converted their 5GHz radio flux
densities to 1.4GHz values using a constant spectral radio index of
$-0.5$}. This value is comparable to the AGN value (0.87, see Eqn.~1),
but that may be coincidental since their sample is so much brighter
than ours (compare the solid line in Fig.~\ref{bestcomp} to our median
values); even though the Wills et al. sample is considered low radio
luminosity, it is still more than 3 orders of magnitude brighter than
our median values. They are probing the brightest part of the radio
source population. This can be put into more perspective by comparing
our sample to the one of \citet{best05}. Their paper is also based on
FIRST and SDSS data. Actually, their data-set is wholly contained
within ours.

In Fig.~\ref{bestcomp}, we replot the data from Fig.~\ref{zandotrend}
onto the data from Best et al. We only include sources from their
sample (open square symbols) that are both in our sample and have
detected [\ion{O}{3}] emission (Fig.~9, lower right panel from Best et
al. includes a set of [\ion{O}{3}] non-detections).  Our median
[\ion{O}{3}] values cover the exact same range as the Best et
al. sample; the main difference is the radio luminosity. Since our
data represent median values, it implies that a full 50\%\ of the
sources are located to the {\it left} of our data-points. The fact
that the Best et al. data can be up to more than 3 orders of magnitude
brighter (in the radio) illustrates the severe radio selection bias of
their sample. It also makes it clear that one has to be careful in
interpreting possible correlations between [\ion{O}{3}] and radio
luminosities. The Best et al. sample would result in a much steeper
relation.

Since our sample has not been selected on radio properties, and since
we include {\it all} sources in the median calculations, we consider
our median [\ion{O}{3}] -- radio luminosity correlation more robust
than previous work based on radio-selected samples\footnote{That is
not to say that our sample is without biases: galaxies in our sample
have to have measurable emission-line ratios.}. Consequently, one can
use the median radio luminosity for the star-formation subsample to
estimate the star-formation rate. \citet{hopkins03} compares various
star-formation indicators using SDSS data, and incorporates the
radio-derived star-formation rate method of \citet{bell03}. The
inferred star-formation rates are listed in Table~3, and range from
less than 1 to about 20 M$_{\odot}$ per year. These modest
star-formation rates are consistent with values derived through, for
instance, H$\alpha$ luminosities of star-forming galaxies that have
been detected by FIRST \citep{hopkins03}. The listed values should not
be taken at face value, however. On the one hand there is an unknown
AGN contribution to the overall radio luminosity which inflates the
star-formation rate, and on the other hand our median stacking
technique is insensitive to extended (beyond $\sim6$ arcseconds) radio
emission, causing it to underestimate the actual radio luminosity and
star-formation rate.

In fact, we measure slightly different sizes for the AGN and
star-formation subsamples: $5.9\arcsec\pm0.3\arcsec$ by
$6.4\arcsec\pm0.4\arcsec$ and $5.2\arcsec\pm0.2\arcsec$ by
$5.7\arcsec\pm0.2\arcsec$, respectively. The AGN sizes are comparable
to the median size for SDSS quasars of 6.4\arcsec\ by 7.0\arcsec\
(also along RA and DEC), as given in Paper~I. This implies that there
is no extended radio emission detected for the star-formation
subsample, while some has been detected for the AGN subsample beyond
the $\sim5.4$\arcsec\ FWHM of the FIRST point-spread-function.

\subsection{Summary}

Combining a large number of objects across a range of AGN activity
from the SDSS with the FIRST radio survey allows us to correlate
various sample statistics with radio emission at levels that are too
low for individual sources to be detected. We are able to measure
median radio emission down to a few tens of $\mu$Jy, well below the
FIRST detection threshold.

We find the following quantities to correlate strongly with the median
radio luminosity:

\begin{enumerate}
\item{D$_{\rm n}$(4000) index / star-formation rate. This index traces
  the amount of star-formation in a given galaxy by measuring the
  amount of blue light relative to the red. Low values of the D$_{\rm
  n}$(4000) index correspond to more active star-formation, and are
  found to have the highest median radio luminosities (more than an
  order of magnitude larger than high-index objects).}
\item{[\ion{O}{3}] luminosity. Higher levels of [\ion{O}{3}]
  luminosity correspond to higher median radio luminosities. This is
  true over at least 3 orders of magnitude. The [\ion{O}{3}] -- radio
  luminosity correlation for the "pure-AGN" subsample is consistent
  with the same relation for quasars, but extended towards lower
  luminosities. The highest levels of [\ion{O}{3}] and median radio
  emission are seen among the "pure-star-formation" subsample, with
  levels comparable to the quasar ones.}
\item{Central velocity dispersion $\sigma$. Generally, higher values
  of $\sigma$ (i.e., larger galaxies) correspond to larger median
  radio luminosities. However, there are clear differences between the
  AGN and star-formation dominated subsamples. For a given median
  radio luminosity, star-forming systems have a much smaller $\sigma$
  than AGN-dominated systems. Conversely, for a given galaxy size and
  $\sigma$, star-formation dominated systems can be an order of
  magnitude more luminous in the radio than the AGN-dominated
  systems.}
\end{enumerate}

We furthermore find that the AGN contribution to the overall median
radio luminosity is roughly constant in these galaxies at the
$5\times10^{27}$ ergs s$^{-1}$ Hz$^{-1}$ level. The other component is
very tightly correlated to the star-formation rate (as derived from
the optical spectrum). This also implies that if one is considering a
system without an AGN, its star-formation rate can be accurately
assessed using the radio luminosity of the system. There is no
indication that this correlation cannot be extended to star-formation
rates below 0.1 M$_{\odot}$ / yr.

It is clear that by studying (median) statistical properties of
carefully selected and representative samples, one can infer subtle
trends that otherwise are either too small to detect, or are masked by
source intrinsic variations. Our median radio luminosity measurements
are at least a few orders of magnitude below typical studies (see
Fig.~\ref{bestcomp}), and provide a much less-biased view of the radio
properties of galaxies. Methods such as that used in this paper hold
great potential when applied to future, large-scale multi-wavelength
surveys.

\acknowledgments

We like to thank the anonymous referee for helpful comments.  The work
by WDV and RHB was partly performed under the auspices of the
U.S. Department of Energy, National Nuclear Security Administration by
the University of California, Lawrence Livermore National Laboratory
under contract No.  W-7405-Eng-48.  JAH acknowledges support of the
GAANN Fellowship from the U.S. department of Education. The authors
also acknowledge support from the National Radio Astronomy
Observatory, the National Science Foundations (grant AST 00-98355),
and the Space Telescope Science Institute.

\renewcommand{\arraystretch}{.6}
\setlength{\voffset}{-20mm}
\setlength{\textheight}{250mm}

\begin{deluxetable}{rccrrr}
\tablewidth{0pt}
\tablenum{1}
\tablecaption{Median Results for Star-formation Binning}
\label{sfrbinned}
\tablehead{
  \colhead{$z_{\rm  median}$} & \colhead{SFR} &
  \colhead{$\log(L_{\rm radio})$} &  \colhead{$F_{\rm peak}$} & \colhead{D$_{\rm n}$(4000)} \\
  \colhead{} & \colhead{[M$_{\odot}$ / yr]} & 
  \colhead{[ergs/s/Hz]} & \colhead{[$\mu$Jy]} & \colhead{}
}
\startdata
0.0784 & 0.15 & 27.69$\pm$0.02 & 61.7$\pm$3.6 & 1.814\\
0.0845 & 0.47 & 27.90$\pm$0.02 & 80.4$\pm$3.6 & 1.754\\
0.0819 & 0.79 & 27.96$\pm$0.02 & 91.0$\pm$3.8 & 1.728\\
0.0832 & 1.12 & 28.06$\pm$0.01 & 101.8$\pm$3.9 & 1.704\\
0.0842 & 1.47 & 28.14$\pm$0.01 & 114.2$\pm$3.4 & 1.697\\
0.0865 & 1.85 & 28.24$\pm$0.01 & 128.0$\pm$3.2 & 1.683\\
0.0908 & 2.26 & 28.26$\pm$0.01 & 127.1$\pm$3.8 &1.671 \\
0.0946 & 2.72 & 28.36$\pm$0.01 & 139.3$\pm$3.6 & 1.653\\
0.0994 & 3.26 & 28.41$\pm$0.01 & 143.9$\pm$3.5 & 1.639\\
0.1045 & 3.90 & 28.54$\pm$0.01 & 172.9$\pm$3.6 & 1.625\\
0.1081 & 4.68 & 28.62$\pm$0.01 & 190.7$\pm$3.8 & 1.612\\
0.1151 & 5.65 & 28.71$\pm$0.01& 193.5$\pm$3.6 & 1.595\\
0.1193 & 6.97 & 28.80$\pm$0.01 & 224.6$\pm$3.4 & 1.584\\
0.1292 & 9.08 & 28.92$\pm$0.01 & 242.9$\pm$3.5 & 1.565\\
0.1481 & 14.1 & 29.15$\pm$0.01 & 316.8$\pm$3.9 & 1.527\\
\enddata
\tablecomments{Both the radio luminosities and flux densities have been 
corrected for the snapshot-bias (see text). The star-formation rate is taken from
Brinchmann et al. (2004)}
\end{deluxetable}

\begin{deluxetable}{ccccccc}
\tablewidth{0pt}
\tablenum{2}
\tablecaption{Results for D$_{\rm n}$(4000)  Binning}
\label{d4kbinned}
\tablehead{
  \colhead{D$_{\rm n}$(4000)} &
  \colhead{$z_{\rm  median}$} & \colhead{$z_{\rm  mean}$} & 
  \colhead{$\log(L_{\rm radio,flux})$} &  \colhead{$\log(L_{\rm radio,lum})$} &  
  \colhead{$\log(L_{\rm [OIII]}/L_\odot)$} & \colhead{$\log(L_{\rm [OII]}/L_\odot)$} \\
  \colhead{} & \colhead{} & \colhead{} & 
  \colhead{[ergs/s/Hz]} & \colhead{[ergs/s/Hz]} &  
  \colhead{} & \colhead{}
}
\startdata
2.044 & 0.0724 & 0.0818 &  28.15 &  28.05 & 6.08 & 6.50\\
1.942 & 0.0826 & 0.0927 &  28.21 &  28.10 & 6.12 & 6.54\\
1.885 & 0.0873 & 0.0963 &  28.20 &  28.05 & 6.18 & 6.57\\
1.836 & 0.0928 & 0.0994 &  28.26 &  28.10 & 6.28 & 6.64\\
1.791 & 0.0951 & 0.1009 &  28.24 &  28.06 & 6.30 & 6.71\\
1.747 & 0.0960 & 0.1027 &  28.30 &  28.14 & 6.39 & 6.79\\
1.705 & 0.0972 & 0.1034 &  28.36 &  28.21 & 6.53 & 6.91\\
1.665 & 0.1008 & 0.1063 &  28.42 &  28.28 & 6.65 & 7.01\\
1.624 & 0.1014 & 0.1077 &  28.48 &  28.35 & 6.79 & 7.17\\
1.584 & 0.1031 & 0.1079 &  28.56 &  28.44 & 6.92 & 7.30\\
1.541 & 0.1053 & 0.1095 &  28.66 &  28.55 & 7.04 & 7.39\\
1.496 & 0.1069 & 0.1106 &  28.73 &  28.62 & 7.17 & 7.53\\
1.446 & 0.1114 & 0.1147 &  28.86 &  28.79 & 7.34 & 7.69\\
1.384 & 0.1143 & 0.1178 &  28.99 &  28.94 & 7.54 & 7.86\\
1.276 & 0.1201 & 0.1244 &  29.21 &  29.19 & 7.85 & 8.11\\
\enddata
\tablecomments{$\log(L_{\rm radio,flux})$ and  $\log(L_{\rm radio,lum})$ are luminosities
based on flux-stacking and luminosity-stacking, respectively. The bolometric luminosity
of the sun ($L_\odot$) is $3.826\times 10^{33}$ ergs s$^{-1}$.}
\end{deluxetable}

\begin{deluxetable}{ccc}
\tablewidth{0pt}
\tablenum{3}
\tablecaption{Inferred Star-formation Rates -- Star-formation subsample}
\label{sfrates}
\tablehead{
  \colhead{$z_{\rm median}$} & \colhead{$\log{(L_{\rm radio})}$} & \colhead{SFR} \\
  \colhead{} & \colhead{[ergs/s/Hz]} & \colhead{[M$_{\odot}$ / yr]}
}
\startdata
0.075 &  27.60 & 0.45 \\
0.095 &  28.10 & 1.07\\
0.100 &  28.34 & 1.61\\
0.101 &  28.46 & 1.97\\
0.106 &  28.57 & 2.37\\
0.108 &  28.72 & 3.06\\
0.112 &  28.78 & 3.38\\
0.115 &  28.91 & 4.49\\
0.119 &  28.99 & 5.40\\
0.123 &  29.08 & 6.64\\
0.128 &  29.23 & 9.38\\
0.134 &  29.34 & 12.1\\
0.149 &  29.65 & 24.7\\
\enddata
\tablecomments{For reference, an $L_*$ galaxy  corresponds to $\log{(L_{\rm radio})} \approx 28.81$ and an SFR of 3.57 M$_{\odot}$ / year.}
\end{deluxetable}

\end{document}